\documentclass[11pt]{article}
\newcommand{\be}{\begin{equation}}
\newcommand{\ee}{\end{equation}}
\newcommand{\bea}{\begin{eqnarray}}
\newcommand{\eea}{\end{eqnarray}}
\newcommand{\sn}{{\rm sn}}

\newcommand{\dn}{{\rm dn}}
\newcommand{\cn}{{\rm cn}}
\newcommand{\sech}{{\rm sech}}

\begin{document}

\vspace{0.5in}
\begin{center}
{\LARGE{\bf Novel Superposed Kinklike and Pulselike Solutions for Several 
Nonlocal Nonlinear Equations}} 
\end{center}

\begin{center}
{\LARGE{\bf Avinash Khare}} \\
{Physics Department, Savitribai Phule Pune University \\
 Pune 411007, India}
\end{center}

\begin{center}
{\LARGE{\bf Avadh Saxena}} \\ 
{Theoretical Division and Center for Nonlinear Studies, 
Los Alamos National Laboratory, Los Alamos, New Mexico 87545, USA}
\end{center}

\vspace{0.9in}
\noindent{\bf {Abstract:}}

We show that a number of nonlocal nonlinear equations including 
the Ablowitz-Musslimani and the Yang variant of the nonlocal nonlinear 
Schr\"od-inger (NLS) equation, nonlocal modified Korteweg de Vries (mKdV) 
equation as well as the nonlocal Hirota equation admit novel kinklike and 
pulselike superposed periodic solutions. Further, we show that the nonlocal 
mKdV equation, in addition also admits the superposed (hyperbolic) 
kink-antikink solution. Besides, we show that while the nonlocal Ablowitz-Musslimani 
variant of the NLS admits complex (parity-time reversal or) PT-invariant kink and 
pulse solutions, neither the local NLS nor the Yang variant of 
the nonlocal NLS admits such solutions. Finally, except for the
Yang variant of the nonlocal NLS, we show that the other three nonlocal 
equations admit both the kink and pulse solutions in the same model.

\section{Introduction}

Few years ago Ablowitz and Musslimani proposed a nonlocal variant of the
nonlinear Schr\"odinger equation (NLS) \cite{abm,abm1} and showed that it is
an integrable system. The interesting point of this nonlocal variant of NLS is that the 
corresponding nonlinearity-induced potential is in general complex and (parity-time 
reversal or) PT-invariant. We note here that in recent years
it has been realized that optics can provide an ideal ground for 
testing some of the consequences of such theories \cite{mus}.
It has been realized that the PT-symmetric optics can give rise to an
entirely new class of optical structures and devices with altogether
new properties \cite{exp}. Few years later, Yang introduced another variant of the 
nonlocal NLS equation \cite{yang18} and showed that it is also an integrable system. 
He obtained it as a special reduction of the Manakov system \cite{manakov}. He has
suggested that his nonlocal NLS variant could be useful in case the
two components of the Manakov system are related by a parity symmetry.
In contrast to the Ablowitz-Musslimani nonlocal NLS, the 
nonlinearity-induced potential in the Yang's nonlocal NLS is real
and symmetric in $x$. In recent years the nonlocal variants of the mKdV \cite{he18} 
and the Hirota equation \cite{ccf,xyx} have also been proposed. 

It is clearly of interest to obtain the various exact solutions admitted
by these nonlocal equations and contrast them with the exact solutions
of the corresponding local equations. Few years back we \cite{ks14} 
considered several nonlocal equations 
including the Ablowitz-Musslimani variant of the nonlocal NLS 
equation \cite{abm} and obtained their exact periodic and 
hyperbolic solutions. One of the main purposes of this paper is to show that 
both the nonlocal variants of NLS, the nonlocal mKdV as well as the nonlocal 
Hirota equation admit periodic {\it superposed} solutions
in terms of $\sn(x+\Delta,m) \pm \sn(x-\Delta,m)$ as well as
$\dn(x+\Delta,m) \pm \dn(x-\Delta,m)$. Here $\sn(x,m)$ and $\dn(x,m)$ are
the Jacobi elliptic functions \cite{as} with $m$ being the modulus 
$0 \le m \le 1$. By superposed we mean that a solution can be expressed  
as a linear combination of two kink solutions or two pulse solutions or the 
corresponding periodic solutions. 

Besides these, we also discuss several other solutions
admitted by these equations. Further, we also discuss those solutions
of the Ablowitz-Musslimani nonlocal variant of the NLS which we had
missed earlier. We show that while the Ablowitz-Musslimani variant
of the nonlocal NLS admits complex PT-invariant kink and pulse solutions,
neither the Yang variant of the nonlocal NLS nor the local NLS 
admits such solutions. Further, while the Ablowitz-Musslimani variant 
of the nonlocal NLS admits both the kink and pulse solutions in
the {\it same} model, neither the Yang variant of the nonlocal NLS, 
nor the local NLS admits these solutions in the same model. 
We also show that while the local mKdV equation is known to admit the 
complex PT-invariant kink and pulse solutions, the corresponding nonlocal 
variant does not admit such solutions. Finally, we show that neither the Hirota 
equation \cite{hirota} nor the nonlocal Hirota equation (discussed in Sec. 5) 
admits complex PT-invariant solutions.

The plan of the paper is the following. In Sec. 2 we show that the 
Yang variant of the nonlocal NLS equation admits several solutions
including novel superposed periodic solutions. In Sec. 3 we consider
the Ablowitz-Musslimani variant of the nonlocal NLS equation and show
that apart from the several solutions already known \cite{ks14}, this
nonlocal equation also admits a few more solutions including the complex
PT-invariant solutions as well as periodic superposed solutions. In Sec. 4 
we consider the nonlocal mKdV equation
and obtain its several new solutions including the periodic as well as
the hyperbolic superposed solutions. In Sec. 5 we consider a nonlocal 
variant of the Hirota equation and obtain its several solutions 
including the periodic superposed solutions. Finally, in Sec. 6 we
summarize the results obtained and point out some of the open
problems. In Appendix A we present many of the exact solutions of 
Yang's nonlocal NLS equation which are also the solution of the local NLS 
equation. In Appendix B we similarly present those solutions of the 
Ablowitz-Musslimani variant of the nonlocal NLS which we had missed in
our earlier paper \cite{ks14} and which are also the solutions of the
local NLS equation. Finally, in Appendix C we present those solutions
of the nonlocal attractive (repulsive) mKdV which are also the 
solutions of the corresponding local attractive (repulsive) mKdV
equation.

\section{Periodic and Hyperbolic Solutions of Yang's version of Nonlocal NLS}

In 2013, Ablowitz and Musslimani introduced a novel nonlocal version of 
the NLS \cite{abm}
\be\label{1}
i\psi_{t}(x,t)+\psi_{xx}(x,t) + 2g \psi^2(x,t) \psi^{*}(-x,t) = 0\,,
~~g = \pm 1\,,
\ee
where the power defined by
\be\label{2}
P = \int_{-\infty}^{\infty} dx\, |\psi(x,t)|^2\,,
\ee
is not conserved but the pseudo-power $Q$ defined by
\be\label{3}
Q = \int_{-\infty}^{\infty} dx\, \psi(x,t) \psi(-x,t)\,,
\ee
is conserved. Here $g = +1$ $(-1)$ corresponds to the attractive (repulsive)
case. In addition, The Hamiltonian $H$ defined by 
\be\label{4}
H = \int_{-\infty}^{\infty} dx\,[\psi_{x}(x,t) \psi^{*}_x(-x,t)
-\frac{g}{2} \psi^{2}(x,t) \psi^{*2}(-x,t)]\,,
\ee
is also conserved. Ablowitz-Musslimani showed that like the usual NLS, 
this nonlocal
NLS is also integrable. Notice that the nonlinearity induced potential
$\psi(x,t) \psi^{*}(-x,t)$ is in general complex but PT-invariant.

Few years later, Yang \cite{yang18} introduced another version of 
the nonlocal NLS given by
\be\label{5}
i\psi_{t}(x,t)+\psi_{xx}(x,t) + g [\psi^2(x,t)+|\psi(-x,t)|^2]\psi(x,t) = 
0\,,~~g = \pm 1\,.
\ee
In this model, not only the power given by Eq. (\ref{2}) is conserved
but also the two different pseudo-powers given by
\be\label{6}
P_1 = \int_{-\infty}^{\infty} dx\, \psi^{*}(-x,t) \psi(x,t)\,,~~
P_2 = \int_{-\infty}^{\infty} dx\, \psi^{*}(x,t) \psi(-x,t)\,,
\ee
are conserved. Besides the Hamiltonian given by
\be\label{7}
H = \int_{-\infty}^{\infty} dx\,[\psi_{x}(x,t) \psi^{*}_x(-x,t)
-\frac{g}{4} [|\psi(x,t)|^2 +|\psi(-x,t)|^2]^{2}\,,
\ee
is also conserved. Further, Yang \cite{yang18} showed that his nonlocal
version of NLS is also integrable. Notice that the nonlinearity-induced
potential $|\psi(x,t)|^2+|\psi(-x,t)|^2$ in Yang's nonlocal NLS is real
and symmetric in $x$. 

We now show that like the Ablowitz-Musslimani nonlocal NLS \cite{ks14}, 
the Yang's nonlocal NLS Eq. (\ref{5}) also admits a large number of 
periodic and hyperbolic solutions. Besides, it also admits novel
superposed periodic solutions. 

{\bf Solution I}

It is easy to show that 
\be\label{8}
\psi(x,t) = A \dn(\beta x, m) e^{i\omega (t +t_0)}\,,
\ee
is an exact solution to Eq. (\ref{5}) provided
\be\label{9}
 g = 1\,,~~A^2 = \beta^2\,,~~\omega = (2-m)\beta^2\,.
\ee
Note that for any solution of the above form, translation shift in time, i.e.
$t_0 \ne 0$ is allowed. For simplicity, from now onwards we put $t_0 = 0$ 
while discussing rest of the solutions, but this point should be kept in
mind. However, unlike the local case, for the nonlocal case, 
a translation shift in $x$ is not allowed. In particular, while 
$A \dn[\beta(x+x_0),m]$ for arbitrary $x_0$ is a solution of the local 
NLS equation in case constraints (\ref{9}) are satisfied, it is the solution
of Eq. (\ref{5}) only in case $x_0 = 0$. 

It turns out that most of the solutions (including the solution as given
by Eq. (\ref{8}))
of the local NLS and the Yang variant of the nonlocal NLS are valid for the
same values of the parameters and we have therefore decided not to 
present them here but for completeness present them in Appendix A. 

As mentioned above, unlike the local NLSE, the solutions of the nonlocal 
NLSE are not invariant with respect to shifts in $x$. For example, while
$A \dn[\beta (x+x_0), m] e^{i\omega t}$ is an exact solution of the local NLSE
no matter what $x_0$ is, it is not an exact solution of the nonlocal Eq. 
(\ref{5}). However, for special values of $x_0$, $\sn(x,m)$, $\cn(x,m)$
and $\dn(x,m)$ are still the solutions of the nonlocal Eq. (\ref{5}).
In particular, we now show that when $x_0 = K(m)$, where $K(m)$ is the 
complete elliptic integral of the first kind, there are exact solutions of 
the nonlocal Eq. (\ref{5}) in both the focusing ($g > 0$) and
the defocusing ($g < 0$) cases. This is because \cite{as}
\bea\label{23}
&&\dn[x+K(m), m] = \frac{\sqrt{1-m}}{\dn(x, m)}\,,~~
\sn[x+K(m), m] = \frac{\cn(x, m)}{\dn(x, m)}\,, \nonumber \\
&&\cn[x+K(m), m] = -\frac{\sqrt{1-m} \sn(x, m)}{\dn(x, m)}\,.
\eea

{\bf Solution II}

It is easy to show that 
\be\label{24}
\psi(x,t) = \frac{A}{\dn(\beta x, m)} e^{i\omega t}\,,
\ee
is an exact solution to Eq. (\ref{5}) provided
\be\label{25}
g=1\,,~~A^2 = (1-m) \beta^2\,,~~\omega = (2-m)\beta^2\,.
\ee

{\bf Solution III}

It is easy to show that 
\be\label{26}
\psi(x,t) = \frac{A\sqrt{m} \sn(\beta x, m)}{\dn(\beta x, m)} 
e^{i\omega t}\,,
\ee
is an exact solution to Eq. (\ref{5}) provided
\be\label{27}
g = 1\,,~~ A^2 = (1-m) \beta^2\,,~~\omega = (2m-1)\beta^2\,.
\ee
It is worth pointing out that in contrast to the Yang variant of the NLS 
Eq. (\ref{5}) or the local NLS, the Ablowitz-Musslimani variant of 
the nonlocal NLS Eq. (\ref{1}) admits the solution (\ref{26}) only
if $g = -1$ \cite{ks14}. 

{\bf Solution IV}

It is easy to show that 
\be\label{28}
\psi(x,t) = \frac{A\sqrt{m} \cn(\beta x, m)}{\dn(\beta x, m)} 
e^{i\omega t}\,,
\ee
is an exact solution to Eq. (\ref{5}) provided $m \ne 1$ and
\be\label{29}
g = -1\,,~~ A^2 = \beta^2\,,~~\omega = -(1+m)\beta^2\,.
\ee

{\bf Solution V}

Remarkably, it turns out that not only $\dn[\beta x,m]$ and 
$\dn[\beta x+K(m),m]$ but even their superposition is a solution
of the nonlocal NLS Eq. (\ref{5}). In particular, it is easy to check
that
\be\label{24a}
\psi(x,t) = e^{i\omega t} \left[A\dn(\beta x,m)+ \frac{B\sqrt{1-m}}
{\dn(\beta x,m)}\right]\,,
\ee
is also an exact solution of the nonlocal Eq. (\ref{5}) provided
\be\label{24b}
g = 1\,,~~ A^2 =  \beta^2\,,~~B = \pm A\,,~~
\omega = [2-m \pm 6\sqrt{1-m}]\beta^2\,,
\ee
where the $\pm$ sign in $B = \pm A$ and in $\omega$ are correlated.

Recently, we \cite{ks22} have shown that the local NLS Equation admits
three novel periodic solutions which can be re-expressed as the superposition
of a periodic kink and an antikink or two periodic pulse solutions. 
Now we show that the nonlocal NLS Eq. (\ref{5}) also admits the same three
novel  periodic solutions which can be re-expressed as superposition of 
a periodic kink and an antikink or two periodic pulse solutions $\sn(x,m)$ 
and $\dn(x,m)$, respectively. Since most of the algebra is the same as in the
local case and is discussed in detail in \cite{ks22}, we avoid giving details here.

{\bf Solution VI}

It is easy to check that the nonlocal NLS Eq. (\ref{5}) 
admits the periodic solution
\be\label{30}
\psi(x,t) = e^{i\omega t} \left[\frac{A \dn(\beta x,m) 
\cn(\beta x,m)}{1+B\cn^2(\beta x,m)}\right]\,,
~~B > 0\,,
\ee
provided $g = -1$ and further 
\bea\label{31}
&&0 < m < 1\,,~~B = +\frac{\sqrt{m}}{1-\sqrt{m}}\,,
\nonumber \\
&&\omega = -[1+m+6\sqrt{m}]\beta^2 < 0\,,~~
A^2 = \frac{4 \sqrt{m} \beta^2}{(1-\sqrt{m})^2}\,.
\eea
Note that this solution is not valid for $m = 1$, i.e. the nonlocal NLS
Eq. (\ref{5}) does not admit a corresponding hyperbolic solution.

At this stage, we recall the well known addition theorem for $\sn(x,m)$ 
\cite{as}, i.e.
\be\label{32}
\sn(a+b,m) = \frac{\sn(a,m) \cn(b,m) \dn(b,m) + \cn(a,m) \dn(a,m) \sn(b,m)}
{1-m \sn^2(a,m) \sn^2(b,m)}\,.
\ee
From here it is straightforward to derive the identity
\be\label{33}
\sn(y+\Delta,m)-\sn(y-\Delta,m) = \frac{2\cn(y,m) \dn(y,m) \frac{\sn(\Delta,m)}
{\dn^2(\Delta,m)}}{1+B\cn^2(y,m)}\,,~~B = \frac{m \sn^2(\Delta,m)}
{\dn^2(\Delta,m)}\,.
\ee
On comparing Eqs. (\ref{30}) and (\ref{33}) and using Eq. (\ref{31}), 
one can re-express the periodic solution (\ref{30}) 
as superposition of a periodic kink and an antikink solution, i.e.
\be\label{34}
\psi(x,t) = e^{i\omega t} \sqrt{\frac{m}{2}} \beta
 [\sn(\beta x +\Delta, m) -\sn(\beta x- \Delta, m)]\,.
\ee
Here $\Delta$ is defined by $\sn(\sqrt{m}\Delta,1/m) = \pm m^{1/4}$,
where use has been made of the identity \cite{as}
\be\label{35}
\sqrt{m} \sn(y,m) = \sn(\sqrt{m} y,1/m)\,,
\ee

{\bf Solution VII}

It is easy to check that the nonlocal NLS Eq. (\ref{5})
admits another periodic solution
\be\label{36}
\psi(x,t) = e^{i\omega t} \frac{A \sn(\beta x,m) 
\cn(\beta x,m)}{1+B\cn^2(\beta x,m)}\,,
\ee
provided
\bea\label{37}
&&0 < m < 1\,,~~B = \frac{1-\sqrt{1-m}}{\sqrt{1-m}}\,,
~~g = 1\,, \nonumber \\
&&\omega = (2-m-6\sqrt{1-m})\beta^2\,,~~ A^2 
= \frac{2(1-\sqrt{1-m})^2 \beta^2}{\sqrt{1-m}}\,.
\eea 

On using the well known addition theorem for $\dn(x,m)$ \cite{as}
\be\label{38}
\dn(a+b,m) = \frac{\dn(a,m) \dn(b,m) - m \sn(a,m) \cn(a,m) \sn(b,m) \cn(b,m)}
{1-m\sn^2(a,m) \sn^2(b,m)}\,,
\ee
one can derive the identity
\be\label{39}
\dn(x-\Delta,m) -\dn(x+\Delta,m) = \frac{2m\sn(\Delta,m) \cn(\Delta,m) 
\sn(x,m) \cn(x,m)} {\dn^2(\Delta,m)[1+\frac{m\sn^2(\Delta,m)}
{\dn^2(\delta,m)} \cn^2(x)]}\,.
\ee
On comparing solutions (\ref{36}) and (\ref{39}) and using 
Eq. (\ref{37}) we find that the solution given in Eq. (\ref{36}) can be 
re-expressed as a superposition of two periodic pulse solutions, i.e.
\be\label{40}
\psi(x,t) = e^{i\omega t}\beta \sqrt{\frac{1}{2}} 
\bigg (\dn[\beta(x)-K(m)/2,m] - \dn[\beta(x)+K(m)/2,m] \bigg )\,.
\ee

{\bf Solution VIII}

The nonlocal NLS Eq. (\ref{5}) also admits another periodic solution
\be\label{41}
\psi(x,t) = e^{i\omega t} 
\frac{A \dn(\beta x,m)}{1+B\cn^2(\beta x,m)}\,,
\ee
provided
\bea\label{42}
&&0 < m < 1\,,~~ B = \frac{1-\sqrt{1-m}}{\sqrt{1-m}}\,, 
~~g = 1\,, \nonumber \\
&&\omega = [2-m+6\sqrt{1-m}]\beta^2\,,
~~A^2 = \frac{4}{\sqrt{1-m}} \beta^2\,.
\eea 
Note that for this solution $g > 0$, $\omega > 0$ and it is not valid for $m = 1$.

Now using the addition theorem for $\dn(x,m)$ as given by Eq. (\ref{38}), 
one can derive another identity, i.e.
\be\label{43}
\dn(x+\Delta,m)+\dn(x-\Delta,m) = \frac{2\dn(x,m)}
{\dn(\Delta,m) (1+B\cn^2(x,m)}\,,~~B = \frac{m \sn^2(\Delta,m)}
{\dn^2(\Delta,m)}\,.
\ee
On comparing Eqs. (\ref{41}) and (\ref{43}) and using Eq. (\ref{42}),
the periodic solution (\ref{41}) can be
re-expressed as superposition of two periodic pulse solutions, i.e.
\be\label{44}
\psi(x,t) = e^{i\omega t} \beta  
[\dn(\beta x +K(m)/2, m)+\dn(\beta x -K(m)/2, m)]\,,
\ee
where $\Delta = \pm K(m)/2$ 

It is worth noting that for both the superposed periodic pulse 
solutions VII and VIII, not only the 
value of $B$ is the same but also $g > 0$ for both the solutions. 
Further, while the solutions IV and VI are valid if $g = -1$, the remaining
six solutions are valid provided $g = 1$.

\section{Superposed Solutions of the Ablowitz-Musslimani version of 
Nonlocal NLS}

Few years ago, we had already obtained \cite{ks14} several exact solutions
of the Ablowitz-Musslimani version of the nonlocal NLS as given by
Eq. (\ref{1}). However, it turns out that we had missed a few exact solutions 
in that paper.  We now mention only those missed solutions which are either 
not the solutions of the local NLS or are not the solutions of the local NLS
for the same values of the parameters. However, for the sake of completeness,
in Appendix B we have given those solutions of the Ablowitz-Musslimani variant
of the nonlocal NLS which are also the solutions of the local NLS as well as
the Yang variant of the nonlocal NLS.

We now show that unlike the local NLS or the Yang version of the nonlocal NLS 
\cite{yang18}, the Ablowitz-Musslimani variant of the
nonlocal NLS admits complex PT-invariant periodic and hyperbolic 
superposed solutions. 

{\bf Solution I}

It is straightforward to show that the nonlocal Eq. (\ref{1}) admits the
complex PT-invariant periodic solution
\be\label{1.4}
\psi(x,t) = e^{i\omega t}[A\dn(\beta x,m) + iB\sqrt{m} \sn(\beta x,m)]\,,
\ee
provided
\be\label{1.5}
B = \pm A\,,~~g = 1\,,~~A = \frac{\beta}{2}\,,~~\omega = 
-\frac{(2m-1)}{2}\beta^2\,.
\ee

{\bf Solution II}

Another complex PT-invariant periodic solution of the nonlocal Eq. (\ref{1}) 
is
\be\label{1.6}
\psi(x,t) = e^{i\omega t}[A\sqrt{m} \cn(\beta x,m) 
+ iB\sqrt{m} \sn(\beta x,m)]\,,
\ee
provided
\be\label{1.7}
B = \pm A\,,~~g = 1\,,~~A = \frac{\beta}{2}\,,~~\omega = 
-\frac{(2-m)}{2}\beta^2\,.
\ee

{\bf Solution III}

In the limit $m = 1$, both the solutions I and II go over to the 
complex PT-invariant hyperbolic solution of the nonlocal Eq. (\ref{1}) 
\be\label{1.8}
\psi(x,t) = e^{i\omega t}[A\sech(\beta x) 
+ iB \tanh(\beta x)]\,,
\ee
provided
\be\label{1.9}
B = \pm A\,,~~g = 1\,,~~A = \frac{\beta}{2}\,,~~\omega = 
-\frac{\beta^2}{2}\,.
\ee

{\bf Solution IV}

Remarkably, the nonlocal Eq. (\ref{1}) also admits the
complex PT-invariant periodic solution
\be\label{1.10}
\psi(x,t) = e^{i\omega t}[A\sqrt{m} \sn(\beta x,m) + iB \dn(\beta x,m)]\,,
\ee
provided the {\it same} relations as given in Eq. (\ref{1.5}) are satisfied. 

{\bf Solution V}

Another complex PT-invariant periodic solution of the nonlocal Eq. (\ref{1}) 
is
\be\label{1.11}
\psi(x,t) = e^{i\omega t}[A\sqrt{m} \sn(\beta x,m) 
+ iB\sqrt{m} \cn(\beta x,m)]\,,
\ee
provided the {\it same} relations as in Eq. (\ref{1.7}) are satisfied.

{\bf Solution VI}

In the limit $m = 1$, both the solutions IV and V go over to the 
complex PT-invariant hyperbolic solution of the nonlocal Eq. (\ref{1}) 
\be\label{1.12}
\psi(x,t) = e^{i\omega t}[A\tanh(\beta x) 
+ iB \sech(\beta x)]\,,
\ee
provided the {\it same} relations as in Eq. (\ref{1.9}) are satisfied.

The Ablowitz-Musslimani variant of the nonlocal NLS is rather unusual in
the sense that the same model (i.e. with $g = 1$) admits not only the
kink and pulse solutions but also the complex PT-invariant pulse and
kink solutions with PT-eigenvalue +$1$ and $-1$, respectively.

We now show that Eq. (\ref{1}) also satisfies four novel periodic 
solutions which can be re-expressed either as the superposition of 
a periodic kink and an antikink or two periodic kink or two
periodic pulse solutions, respectively.

{\bf Solution VII}

It is easy to check that the nonlocal NLS Eq. (\ref{1}) 
admits the periodic solution
\be\label{1.13}
\psi(x,t) = e^{i\omega t} \left[\frac{A \dn(\beta x,m) 
\cn(\beta x,m)}{1+B\cn^2(\beta x,m)}\right]\,,
~~B > 0\,,
\ee
provided $g = -1$ and further 
\bea\label{1.14}
&&0 < m < 1\,,~~B = \frac{\sqrt{m}}{1-\sqrt{m}}\,,
\nonumber \\
&&\omega = -[1+m+6\sqrt{m}]\beta^2 < 0\,,~~
A^2 = \frac{4 \sqrt{m} \beta^2}{(1-\sqrt{m})^2}\,.
\eea
Note that this solution is not valid for $m = 1$, i.e. the nonlocal NLS
Eq. (\ref{1}) does not admit a corresponding hyperbolic solution.

On comparing Eqs. (\ref{1.13}) and the identity (\ref{33}) and 
using Eq. (\ref{1.14}), 
one can re-express the periodic solution (\ref{1.13}) 
as superposition of a periodic kink and a periodic antikink solution, i.e.
\be\label{1.15}
\psi(x,t) = e^{i\omega t} \sqrt{\frac{m}{2}} \beta
 [\sn(\beta x +\Delta, m) -\sn(\beta x- \Delta, m)]\,. 
\ee
Here $\Delta$ is defined by $\sn(\sqrt{m}\Delta,1/m) = \pm m^{1/4}$,
where use has been made of the identity (\ref{35}).

{\bf Solution VIII}

Remarkably, Eq. (\ref{1}) also admits another periodic solution
\be\label{1.16}
\psi(x,t) = e^{i\omega t} \frac{A \sn(\beta x,m)}{1+B\cn^2(\beta x,m)}\,,
~~B > 0\,,
\ee
provided
\bea\label{1.17}
&&0 < m < 1\,,~~B = \frac{\sqrt{m}}{1-\sqrt{m}}\,,~~g = -1\,,
\nonumber \\
&&a = [6\sqrt{m}-(1+m)]\beta^2\,,~~A^2 = 4\sqrt{m} \beta^2\,.
\eea 
Note that this solution too is not valid in the hyperbolic limit of 
$m = 1$. Now on using the $\sn(x,m)$ addition theorem (\ref{32}), one can 
derive another identity
\be\label{1.18}
\sn(y+\Delta,m)+\sn(y-\Delta,m) = \frac{2\sn(y,m) \frac{\cn(\Delta,m)}
{\dn(\Delta,m)}}{1+B\cn^2(y,m)}\,,~~B = \frac{m \sn^2(\Delta,m)}
{\dn^2(\Delta,m)}\,.
\ee
On comparing Eqs. (\ref{1.16}) and (\ref{1.18}) and using Eq. (\ref{1.17}), 
the periodic solution VIII given by Eq. (\ref{1.16}) can be
re-expressed as superposition of two periodic kink solutions
\be\label{1.19}
\psi(x,t) = i e^{i\omega t} \sqrt{m} \beta  
\bigg [\sn(\beta x +\Delta, m)+\sn(\beta x -\Delta, m) \bigg ]\,. 
\ee
Here $\Delta$ is defined by $\sn(\sqrt{m}\Delta,1/m) = \pm m^{1/4}$,
where use has been made of the identity (\ref{35}).

It is worth noting that for both the solutions VII and VIII, not only 
the value of $B$ is the same but even $g < 0$ for both the solutions.

{\bf Solution IX}

It is easy to check that the nonlocal NLS Eq. (\ref{1})
admits another periodic solution
\be\label{1.20}
\psi(x,t) = e^{i(\omega t)} \frac{A \sn(\beta x,m) 
\cn(\beta x,m)}{1+B\cn^2(\beta x,m)}\,,
\ee
provided
\bea\label{1.21}
&&0 < m < 1\,,~~B = \frac{1-\sqrt{1-m}}{\sqrt{1-m}}\,,
~~g = -1\,, \nonumber \\
&&\omega = (2-m-6\sqrt{1-m})\beta^2\,,~~ A^2 
= \frac{2(1-\sqrt{1-m})^2 \beta^2}{\sqrt{1-m}}\,.
\eea 
Note that whereas this solution is valid if $g = -1$, the same
solution in the Yang's nonlocal case, as well as in the local NLS case 
is valid only if $g = +1$. 

On comparing the solution (\ref{1.20}) with the identity (\ref{39})
and using Eq. (\ref{1.21}) we find that the solution 
as given by Eq. (\ref{1.20}) can be re-expressed as a 
superposition of two periodic pulse solutions, i.e.
\be\label{1.24}
\psi(x,t) = e^{i\omega t}\beta \sqrt{\frac{1}{2}} 
\left(\dn[\beta(x)-K(m)/2,m] - \dn[\beta(x)+K(m)/2,m] \right)\,.
\ee

{\bf Solution X}

Remarkably, the nonlocal NLS Eq. (\ref{1}) also 
admits another periodic solution
\be\label{1.25}
\psi(x,t) = e^{i\omega t} 
\frac{A \dn(\beta x,m)}{1+B\cn^2(\beta x,m)}\,,
\ee
provided
\bea\label{1.26}
&&0 < m < 1\,,~~ B = \frac{1-\sqrt{1-m}}{\sqrt{1-m}}\,, 
~~g = 1\,, \nonumber \\
&&\omega = [2-m+6\sqrt{1-m}]\beta^2\,,
~~A^2 = \frac{4}{\sqrt{1-m}} \beta^2\,.
\eea 
Note that this solution is also not valid for $m = 1$.
Thus for this solution $g > 0$, $\omega > 0$.

On comparing the solution (\ref{1.25}) and the identity (\ref{43}) 
and using Eq. (\ref{1.26}), the periodic solution (\ref{1.25}) can be
re-expressed as superposition of two periodic pulse solutions, i.e.
\be\label{1.27}
\psi(x,t) = e^{i\omega t} \beta 
[\dn(\beta x +K(m)/2, m)+\dn(\beta x -K(m)/2, m)]\,,
\ee
where $\Delta = \pm K(m)/2$ 

It is worth noting that for both the superposed periodic pulse 
solutions IX and X, while the value of $B$ is the same but the value 
of $g$ is opposite for the two solutions. 

Note that out of the 10 solutions, the solutions VII to X are only
valid if $m \ne 1$. Further, while the solutions VII, VIII and IX are 
valid if $g < 0$, the remaining seven solutions are valid if $g > 0$. 

\section{Periodic and Hyperbolic Solutions of Nonlocal mKdV Equation}

Recently, a nonlocal variant of the mKdV equation (both 
attractive and repulsive) has been proposed \cite{he18} and is 
given by
\be\label{2.1}
\psi_t(x,t) + \psi_{xxx}(x,t) + 6g \psi(x,t) \psi(-x,-t) \psi_{x}(x,t) = 0\,,
\ee
where $g = 1$ $(-1$) corresponds to attractive (repulsive) nonlocal mKdV.
Let us first note that all those solutions of the attractive (repulsive) local 
MKdV equation
\be\label{2.2}
\psi_t(x,t) + \psi_{xxx}(x,t) +6g \psi^2(x,t) \psi_{x}(x,t) = 0\,,
\ee
for which $\psi(-x,-t) = \psi(x,t)$ are obviously also the solutions of the 
corresponding nonlocal attractive (repulsive) mKdV Eq. (\ref{2.1}).
However, because of the translational invariance, unlike the local mKdV
Eq. (\ref{2.2}), the nonlocal mKdV Eq. (\ref{2.1}) does not admit arbitrary
translational shift in $x$ or $t$.

{\bf Solution I}

For example, one of the exact solutions of the nonlocal mKdV 
Eq. (\ref{2.1}) as well as the local mKdV Eq. (\ref{2.2}) is
\be\label{2.4}
\psi(x,t) = A \dn(\xi, m) \,,~~\xi = \beta(x-vt)\,,
\ee
provided
\be\label{2.5}
g = 1\,,~~A^2 = \beta^2\,,~~v = (2-m)\beta^2\,.
\ee
Hence we do not present those solutions of the nonlocal mKdV Eq. (\ref{2.1})
for which $\psi(-x,-t) = \psi(x,t)$ here. However, for the sake of 
completeness, we have given those solutions in Appendix C. 

One comment is in order here. Because of the translational invariance, even 
$\psi(x,t) = A \dn(\xi+\xi_0, m)$ is
an exact solution of the corresponding local attractive mKdV Eq. (\ref{2.2})
under the {\it same} conditions (\ref{2.5}). Here $\xi_0$ is an arbitrary 
constant. The major difference between the local and the nonlocal case is 
that in the nonlocal case neither the translation in $x$ nor $t$ and hence 
$\xi = \beta(x-vt)$ is allowed and the solution can only exist
provided $\xi = 0$. This remark is valid for all the solutions mentioned below, 
for both the attractive as well as the repulsive mKdV Eq. (\ref{2.1}) 
and hence we will not repeat this comment now onwards.

{\bf Solution II}

Remarkably, unlike the local mKdV, for the nonlocal case even 
\be\label{2.12}
\psi(x,t) = A \sqrt{m} \sn(\xi, m) \,,~~\xi = \beta(x-vt)
\ee
is an exact solution to the attractive mKdV, i.e. of Eq. (\ref{2.1}) provided
\be\label{2.13}
g = 1\,,~~A^2 = \beta^2\,,~~v = -(1+m)\beta^2\,.
\ee
Note that the local mKdV Eq. (\ref{2.2}) admits this solution only if 
$g = -1$.

{\bf Solution III}

In the limit $m = 1$, the solution II goes over to the hyperbolic solution
\be\label{2.14}
\psi(x,t) = A \tanh(\xi) \,,~~\xi = \beta(x-vt)\,,
\ee
provided
\be\label{2.15}
g = 1\,,~~A^2 = \beta^2\,,~~v = -2\beta^2\,.
\ee
Thus unlike the local mKdV, the same nonlocal attractive mKdV Eq. (\ref{2.1}) 
(i.e. with $g = +1$) admits both the kink and pulse solutions. This is
because for both the solutions II and III, $\psi(-x,-t) = -\psi(x,t)$.

As mentioned above, unlike the local case, the 
solutions of the nonlocal mKdV Eq. (\ref{2.1})
 are not invariant with respect to shifts in $x$ and $t$. 
However, for special values of $\xi = \beta(x-vt)$, $\sn(x,m)$, $\cn(x,m)$
and $\dn(x,m)$ are still the solutions of the nonlocal Eq. (\ref{2.1}).
In particular, we now show that when the shift is by $K(m)$, where $K(m)$ 
is the complete elliptic integral of the first kind, there are exact solutions of 
the nonlocal Eq. (\ref{2.1}) in both the focusing ($g > 0$) and
the defocusing ($g < 0$) cases. This is because of the relations 
(\ref{23}).

{\bf Solution IV}

It is easy to show that 
\be\label{2.16}
\psi(x,t) = \frac{A}{\dn(\xi, m)}\,,~~\xi = \beta(x-vt)\,,
\ee
is an exact solution to Eq. (\ref{2.1}) (as well as to the local
mKdV Eq. (\ref{2.2})) provided
\be\label{2.17}
g = 1\,,~~A^2 = (1-m) \beta^2\,,~~v = (2-m)\beta^2\,.
\ee

{\bf Solution V}

It is easy to show that 
\be\label{2.17a}
\psi(x,t) = \frac{A\sqrt{m} \sn(\xi, m)}{\dn(\xi, m)}\,,
~~\xi = \beta(x-vt)\,,
\ee
is an exact solution to the nonlocal mKdV Eq. (\ref{2.1}) provided
\be\label{2.18a}
g = -1\,,~~A^2 = (1-m) \beta^2\,,~~v = (2m-1)\beta^2\,.
\ee

{\bf Solution VI}

It is easy to show that 
\be\label{2.18}
\psi(x,t) = \frac{A\sqrt{m} \cn[\beta(x-vt), m]}{\dn[\beta (x-vt), m]}\,,
\ee
is an exact solution to Eq. (\ref{2.1}) provided
\be\label{2.19}
g = -1\,,~~A^2 = \beta^2\,,~~v = -(1+m)\beta^2\,,~~m \ne 1\,.
\ee

{\bf Solution VII}

Remarkably, it turns out that not only $\dn[\beta(x-vt),m]$ and 
$\dn[\beta(x-vt)+K[m]),m]$ but even their superposition is a solution
of the nonlocal mKdV Eq. (\ref{2.1}). In particular, it is easy to check
that
\be\label{2.16a}
\psi(x,t) = A\dn[\beta(x-vt),m]+ \frac{B\sqrt{1-m}}
{\dn[\beta(x-vt),m]}\,,
\ee
is also an exact solution of the nonlocal Eq. (\ref{5}) provided
\be\label{2.16b}
g=1\,,~~ A^2 =  \beta^2\,,~~B = \pm A\,,~~v = [2-m \pm 6\sqrt{1-m}]\beta^2\,,
\ee
where the $\pm$ sign in $B = \pm A$ and in $v$ are correlated.

{\bf Solution VIII}

It has been shown in \cite{kksh} that 
\be\label{2.27}
\psi(x,t) = -2\frac{d}{dx} \tanh^{-1} \bigg [\alpha \sn(ax+bt, k)
\sn(cx+dt, m) \bigg ]\,,
\ee
is an exact solution of the local mKdV Eq. (\ref{2.2}) in case $g = -1$. 
Since for this
solution $\psi(-x,-t) = -\psi(x,t)$, hence it is clear that Eq. (\ref{2.27})
is an exact solution of the nonlocal mKdV Eq. (\ref{2.1}) provided
\bea\label{2.28}
&&g = 1\,,~~k a^{4} = m c^4\,,~~\alpha^2 = \sqrt{km}\,, \nonumber \\
&&b = a[a^2(1+k)+3c^2(1+m)]\,,~~d = c[3a^2(1+k)+(1+m)c^2]\,.
\eea

{\bf Solution IX}

It has been shown in \cite{kksh} that there is a periodic solution of the
local mKdV Equation (\ref{2.2}) 
\be\label{2.30}
\psi(x,t) = -2\frac{d}{dx} \tan^{-1} \bigg [\alpha \cn(ax+bt+a_0, k)
\cn(cx+dt+c_0, m) \bigg ]\,,
\ee
provided
\bea\label{2.31}
&&g = 1\,,~~k(1-k) a^{4} = m(1-m) c^4\,,~~\alpha^2 = \frac{km}{(1-k)(1-m)}\,, 
\nonumber \\
&&b = a[a^2(1-2k)+3c^2(1-2m)]\,,~~d = c[3a^2(1-2k)+(1-2m)c^2]\,.
\eea

Now observe that for this solution $\psi(-x,-t) = -\psi(x,t)$ provided $a_0 
= c_0 = 0$ and hence
\be\label{2.32}
\psi(x,t) = -2\frac{d}{dx} \tan^{-1} \bigg [\alpha \cn(ax+bt, k)
\cn(cx+dt, m) \bigg ]\,,
\ee
is in fact a solution of the nonlocal mKdV Eq. (\ref{2.1}) provided the 
constraints (\ref{2.31}) are satisfied except that $g = -1$.

Recently, we \cite{ks22} have obtained four novel periodic and one hyperbolic 
solutions of the local mKdV Eq. (\ref{2.2}) and shown that they can be 
re-expressed as a superposed kink or pulse solution. We now show that the 
nonlocal mKdV Eq. (\ref{2.1}) also admits these five 
superposed solutions.

{\bf Solution X}

Following \cite{ks22} it is easy to see that
\be\label{2.33}
\psi(x,t) = \frac{A \dn(\xi,m) \cn(\xi,m)}{1+B\cn^2(\xi,m)}\,,~~B > 0\,,
~~\xi = \beta(x-vt)\,,
\ee
is an exact solution of the nonlocal mKdV Eq. (\ref{2.1}) provided
\bea\label{2.34}
&&g = -1\,,~~0 < m < 1\,,~~B = \frac{\sqrt{m}}{1-\sqrt{m}}\,,
\nonumber \\
&&v = -[1+m+6\sqrt{m}]\beta^2 < 0\,,
~~A^2 = \frac{4 \sqrt{m} \beta^2}{(1-\sqrt{m})^2}\,.
\eea 
Note that this solution is not valid for $m = 1$, i.e. the MKdV 
Eq. (\ref{2.1}) does not admit a corresponding hyperbolic solution.
Notice that for this solution $v < 0$.

On using the identity (\ref{33}) one can then rewrite the periodic 
pulse solution (\ref{2.33}) as the superposition of a periodic kink 
and a periodic antikink solution, i.e. 
\be\label{2.35}
\psi(x,t) = \sqrt{2m} \beta 
[\sn(\xi +\Delta, m) -\sn(\xi- \Delta, m)]\,,~~\xi = \beta(x-vt)\,.
\ee
Here $\Delta$ is defined by $\sn(\sqrt{m}\Delta,1/m) = \pm m^{1/4}$,
where use has been made of the identity (\ref{35}). 

{\bf Solution XI}

Following \cite{ks22} it is easy to show that the nonlocal mKdV 
Eq. (\ref{2.1}) admits the periodic kink solution
\be\label{2.36}
\psi(x,t) = \frac{A \sn(\xi,m)}{1+B\cn^2(\xi,m)}\,,~~B > 0\,,
\ee
provided
\bea\label{2.37}
&&g = -1\,,~~0 < m < 1\,,~~B = \frac{\sqrt{m}}{1-\sqrt{m}}\,,
\nonumber \\
&&v = [6\sqrt{m}-(1+m)]\beta^2\,,~~ A^2 = 4\sqrt{m} \beta^2\,.
\eea 
Note that this solution does not exist for $m = 1$, i.e. the corresponding
hyperbolic solution does not exist. 
Notice that for this solution
$\psi(-x,-t) = -\psi(x,t)$ and hence unlike the local mKdV, the nonlocal mKdV
Eq. (\ref{2.1}) admits such a solution in the repulsive case ($g = -1$).

On using the identity (\ref{1.18}), the periodic solution (\ref{2.36}) 
can be re-expressed as
\be\label{2.38}
\psi(x,t) = i \sqrt{m} \beta  
[\sn(\xi +\Delta, m)+\sn(\xi -\Delta, m)]\,,~~\xi = \beta(x-vt)\,.
\ee
Here $\Delta$ is defined by $\sn(\sqrt{m}\Delta,1/m) = \pm m^{1/4}$,
where use has been made of the identity (\ref{35}).

{\bf Solution XII}

Following \cite{ks22} it is easy to show that another 
periodic solution 
to the nonlocal mKdV Eq. (\ref{2.1}) is
\be\label{2.39}
\psi(x,t) = \frac{A \sn(\xi,m) \cn(\xi,m)}{1+B\cn^2(\xi,m)}\,,
\ee
provided
\bea\label{2.40}
&&0 < m < 1\,,~~B = \frac{1-\sqrt{1-m}}{\sqrt{1-m}}\,,~~
v = (2-m-6\sqrt{1-m})\beta^2\,, \nonumber \\
&&g = -1\,,~~A^2 = \frac{4(1-\sqrt{1-m})^2 \beta^2}{\sqrt{1-m}}\,.
\eea
Notice that for this solution
$\psi(-x,-t) = -\psi(x,t)$ and hence unlike the local mKdV, the nonlocal mKdV
Eq. (\ref{2.1}) admits such a solution in the repulsive case ($g = -1$).

On using the identity (\ref{39}), the periodic solution (\ref{2.39})  
can be re-expressed as a 
superposition of two periodic pulse solutions, i.e.
\be\label{2.41}
\psi(x,t) = \beta \big (\dn[\xi -K(m)/2,m] 
- \dn[\xi +K(m)/2,m] \big )\,, ~~\xi = \beta(x-vt)\,.
\ee

{\bf Solution XIII}

Following \cite{ks22}, yet another periodic solution to the nonlocal mKdV 
Eq. (\ref{2.1}) is
\be\label{2.42}
\psi(x,t) = \frac{A \dn(\xi,m)}{1+B\cn^2(\xi,m)}\,,
\ee
provided
\bea\label{2.43}
&&0 < m < 1\,,~~ B = \frac{1-\sqrt{1-m}}{\sqrt{1-m}}\,,~~g = 1\,,
\nonumber \\
&&v = [2-m+6\sqrt{1-m}]\beta^2\,,~~A^2 = \frac{4}{\sqrt{1-m}} \beta^2\,.
\eea
Thus for this solution $v > 0$.

On using the identity (\ref{43}) the periodic solution (\ref{2.42}) can be
re-expressed as superposition of two periodic pulse solutions, i.e.
\be\label{2.44}
\psi(x,t) = \beta  
[\dn(\beta x +K(m)/2, m)+\dn(\beta x -K(m)/2, m)]\,.
\ee

Summarizing, while the solutions X and XII are satisfied for the same
values of the parameters in both the local and nonlocal mKdV Eqs. (\ref{2.2})
and (\ref{2.1}) respectively, the solutions XI and XII are satisfied 
for the opposite values of $g$ in the two models.

{\bf Superposed Solution XIV}

Following \cite{ks22, tan} it is easy to check that the repulsive
mKdV Eq. (\ref{2.1}) admits a hyperbolic pulse solution
\be\label{2.45}
\psi(x,t) = 1-\frac{A}{B+\cosh^2(\xi)}\,,~~B > 0\,,~~\xi = \beta(x-vt)\,,
\ee
provided 
\be\label{2.46}
g = -1\,,~~A = 2\sqrt{B(B+1)}\beta\,,~~\beta^2 = \frac{4(B+1)}{(B+2)^2} < 1\,,~~
v = 4\beta^2 -6\,.
\ee
Now by starting from the identity (\ref{33}) and taking $m = 1$, we obtain
the corresponding hyperbolic identity
\be\label{2.47}
\tanh(y+\Delta)-\tanh(y-\Delta) = \frac{\sinh(2\Delta)}{B+\cosh^2(y)}\,,~~
B = \sinh^2(\Delta)\,.
\ee
On using the identity (\ref{2.47}), the solution (\ref{2.45}) can be
re-expressed as the superposition of a kink and an antikink solution
\be\label{2.48}
\psi(x,t) = 1-\beta[\tanh(\xi+\Delta) -\tanh(\xi-\Delta)]\,,
\ee
where $\xi = \beta(x-vt)$ while $\sinh(\Delta) = \sqrt{B}$.

\section{Exact Solutions of Nonlocal Hirota Equation}

Recently it has been proposed \cite{ccf,xyx} that the nonlocal Hirota 
equation is given by
\bea\label{3.1}
&&iu_t(x,t) +\alpha[u_{xx}(x,t)+2g u^{2}(x,t) u(-x,-t)] \nonumber \\
&&+i\beta[u_{xxx}(x,t) +6g u(x,t) u(-x,-t) u_x(x,t)] = 0\,.
\eea
Note that in case $\beta = 0$ we have an attractive or repulsive nonlocal NLS
depending on whether $g = 1$ or $g = -1$, respectively.
Similarly, when $\alpha = 0$, and $u$ is real we have an 
attractive or repulsive nonlocal  mKdV depending on whether $g = 1$ or $g = -1$, 
respectively. Note that in case $u(-x,-t) = u(x,t)$, then all
the solutions of the local Hirota Equation
\be\label{3.2}
iu_t +\alpha[u_{xx}+2g|u|^2 u]+i\beta[u_{xxx}+6g |u|^2 u_x] = 0\,,
\ee
are also the solutions of the nonlocal Hirota Eq. (\ref{3.1}). 
On the other hand, in case $u(-x,-t) = -u(x,t)$, then all the solutions of 
$g = 1$ $(-1)$ Hirota Eq. (\ref{3.2}) are also the solutions of the nonlocal 
Hirota Eq. (\ref{3.1}) in case $g = -1$ $(+1)$. 

We start from the local Hirota Eq. (\ref{3.2}) and choose the ansatz
\be\label{3.3}
u(x,t) = e^{i\omega t} \phi(\xi)\,,~~\xi = x-vt\,,
\ee
where $\phi$ is real. 
On substituting the ansatz (\ref{3.3}) in Eq. (\ref{3.2}) we obtain
\bea\label{3.4}  
\alpha[\phi_{\xi \xi}+2g\phi^3-\frac{\omega}{\alpha}\phi]  
+i\beta \frac{d}{d\xi}[\phi_{\xi \xi} +2g\phi^3-\frac{v\phi}{\beta}] 
= 0\,.
\eea
It then follows that all the solutions of the real part of 
Eq. (\ref{3.4}), i.e. 
\be\label{3.5}
\phi_{\xi \xi} = \frac{\omega}{\alpha}\phi - 2g \phi^3\,,
\ee
are automatically also the solutions of the imaginary part of
Eq. (\ref{3.4}) provided
\be\label{3.6}
\omega \beta = v \alpha\,.
\ee

It is then straightforward to obtain the solutions of Eq. 
(\ref{3.5}) and hence the solutions of the nonlocal Eq. 
(\ref{3.1}) which we now mention one by one.

{\bf Solution I}

One of the periodic pulse-like solutions of the nonlocal (as well as 
the local) Hirota equation (\ref{3.1})  is
\be\label{3.7}
u(x,t) = A e^{i\omega t} \cn(\xi,m)\,,~~\xi = \delta(x-vt)\,,
\ee
provided the relation (\ref{3.6}) is satisfied and further
\be\label{3.8}
g=1\,,~~A^2 = m \delta^2\,,~~\omega = (2m-1)\alpha \delta^2\,.
\ee

{\bf Solution II}

Another periodic pulse-like solution of the nonlocal (as well as 
the local) Hirota Eq. (\ref{3.1}) is
\be\label{3.9}
u(x,t) = A e^{i\omega t} \dn(\xi,m)\,,
\ee
provided the relation (\ref{3.6}) is satisfied and further if 
\be\label{3.10}
g = 1\,,~~A^2 = \beta^2\,,~~\omega = (2-m) \alpha \delta^2\,.
\ee

{\bf Solution III}

In the limit $m = 1$, both the solutions I and II go over 
to the hyperbolic pulse-like solution
\be\label{3.11}
u(x,t) = A e^{i\omega t} \sech(\xi)\,,
\ee
provided the relation (\ref{3.6}) is satisfied and further if
\be\label{3.12}
g = 1\,,~~A^2 = \beta^2\,,~~\omega = \alpha \delta^2\,.
\ee

{\bf Solution IV}

The nonlocal Hirota Eq. (\ref{3.1}) also admits the periodic 
kink-like solution
\be\label{3.13}
u(x,t) = A e^{i\omega t} \sn(\xi,m)\,,
\ee
provided the relation (\ref{3.6}) is satisfied and further if
\be\label{3.14}
g = 1\,,~~A^2 = m\beta^2\,,~~\omega = -(1+m) \alpha \delta^2\,. 
\ee
Note that in contrast, for the local Hirota Eq. (\ref{3.2}), 
(\ref{3.13}) is a solution only if $g = -1$.

{\bf Solution V}

In the limit $m = 1$, the periodic kink-like  solution (\ref{3.13}) 
goes over to the celebrated single kink-like solution
\be\label{3.15}
u(x,t) = A e^{i\omega t} \tanh(\xi)\,,
\ee
provided the relation (\ref{3.6}) is satisfied and if further
\be\label{3.16}
g = 1\,,~~A^2 = \beta^2\,,~~\omega = -2 \alpha \delta^2\,.
\ee
Note that unlike the Hirota Eq. (\ref{3.2}), the nonlocal Hirota 
Eq. (\ref{3.1}) admits both the kink and pulse solutions in the
same model (i.e. with $g = 1$). 

{\bf Solution VI}

Remarkably, the nonlocal (as well as the local) Hirota Eq. (\ref{3.1}) 
also admits superposition 
of the two periodic pulse-like solutions, i.e.
\be\label{3.17}
u(x,t) =  e^{i\omega t} [A \dn(\xi,m)+ B\sqrt{m} \cn(\xi,m)]\,,
\ee
provided the relation (\ref{3.6}) is satisfied and further if
\be\label{3.18}
g = 1\,,~~B^2 = A^2\,,~~4A^2 = \beta^2\,,~~\omega = (1+m)/2 \alpha \delta^2\,.
\ee

Note that in the limit $m = 1$ and $B = A$, the solution VI also goes over
to the pulse solution III while the solution with $B = -A$ goes to the
vacuum solution $u = 0$.

As mentioned above for the nonlocal NLS and nonlocal mKdV equations, even for 
the Hirota equation, unlike the local case, the solutions of the nonlocal 
Hirota Eq. (\ref{3.1}) are not invariant with respect to shifts in $x$, 
$t$ and hence $\xi$. For example, while
$A \dn[\xi + x_0, m] e^{i\omega t}$ is an exact solution of the 
local Hirota Eq. (\ref{3.2})
no matter what $x_0$ is, it is not an exact solution of the nonlocal Eq. 
(\ref{3.1}). However, for special values of $x_0$, $\sn(x,m)$, $\cn(x,m)$
and $\dn(x,m)$ are still the solutions of the nonlocal Hirota Eq. (\ref{3.1}).
In particular, we now show that when $x_0 = K(m)$, where $K(m)$ is the 
complete elliptic integral of the first kind, there are exact solutions of 
the nonlocal Hirota Eq. (\ref{3.1}) in both the focusing ($g > 0$) and
defocusing ($g < 0$) cases. This is because \cite{as} of the relations
(\ref{23}).

{\bf Solution VII}

Yet another periodic solution to the nonlocal Hirota Eq. (\ref{3.1}) is
\be\label{3.19}
u(x,t) = e^{i\omega t} \frac{A}{\dn(\xi,m)}\,,
\ee
provided the relation (\ref{3.6}) is satisfied and further if
\be\label{3.20}
g = 1\,,~~A^2 = (1-m)\beta^2\,,~~\omega = (2-m) \alpha \delta^2\,.
\ee

{\bf Solution VIII}

Yet another periodic solution to the nonlocal Hirota Eq. (\ref{3.1}) is
\be\label{3.21}
u(x,t) = e^{i\omega t} \frac{A\sqrt{m} \sn(\xi,m)}{\dn(\xi,m)}\,,
\ee
provided the relation (\ref{3.6}) is satisfied and further if
\be\label{3.22}
g = -1\,,~~A^2 = (1-m)\beta^2\,,~~\omega = (2-m) \alpha \delta^2\,.
\ee

{\bf Solution IX}

Yet another periodic solution to the nonlocal Hirota Eq. (\ref{3.1}) is
\be\label{3.23}
u(x,t) = e^{i\omega t} \frac{A\sqrt{m} \cn(\xi,m)}{\dn(\xi,m)}\,,
\ee
provided the relation (\ref{3.6}) is satisfied and further if
\be\label{3.24}
g = -1\,,~~0 < m < 1\,,~~A^2 = \beta^2\,,~~\omega = -(1+m)/2 \alpha \delta^2\,.
\ee

{\bf Solution X}

Remarkably, it turns out that not only $\dn[\xi,m]$ and 
$\dn[\xi+K(m)),m]$ but even their superposition is a solution
of the nonlocal Hirota Eq. (\ref{3.1}). In particular, it is easy to check
that
\be\label{3.19a}
\psi(x,t) = e^{i\omega t} \left[A\dn(\xi,m)+ \frac{B\sqrt{1-m}}
{\dn(\xi,m)}\right]\,,
\ee
is also an exact solution of the nonlocal Eq. (\ref{3.1}) provided
\be\label{3.19b}
g =1\,,~~ A^2 =  \beta^2\,,~~B = \pm A\,,~~\omega 
= [2-m \pm 6\sqrt{1-m}]\beta^2\,,
\ee
where the $\pm$ sign in $B = \pm A$ and in $\omega$ are correlated.

We now mention three periodic solutions of the Hirota Eq. (\ref{1}) which 
can be written as the superposition of either the
periodic kink or pulse solutions $\sn(x,m)$ or $\dn(x,m)$, respectively.

{\bf Solution XI}

One of the periodic solutions of the nonlocal Hirota Eq. (\ref{3.1}) is
\be\label{3.25}
u(x,t) = e^{i\omega t} \frac{A \dn(\xi,m) \cn(\xi,m)}
{1+B\cn^2(\xi,m)}\,,~~B > 0\,,
\ee
provided the relation (\ref{3.6}) is satisfied and further if
\bea\label{3.26}
&&B = \frac{\sqrt{m}}{1-\sqrt{m}}\,,~~g = -1\,,~~
\omega = -[1+m+6\sqrt{m}]\alpha \delta^2 < 0\,, \nonumber \\
&&0 < m < 1\,,~~A^2 = \frac{4 \sqrt{m} \beta^2}{(1-\sqrt{m})^2}\,.
\eea

On using the identity (\ref{33}), the solution (\ref{3.25}) can be re-expressed as 
a superposition of a periodic kink and a periodic antikink solution, i.e.
\be\label{3.27}
u(x,t) = e^{i\omega t} \sqrt{m} \beta 
[\sn(\xi +\Delta, m) -\sn(\xi - \Delta, m)]\,.
\ee
Here $\Delta$ is defined by $\sn(\sqrt{m}\Delta,1/m) = \pm m^{1/4}$,
where use has been made of the identity (\ref{35}).

{\bf Solution XII}

Another periodic solution to the nonlocal Hirota Eq. (\ref{3.1}) is
\be\label{3.28}
u(x,t) = e^{i\omega t} \frac{A \sn(\xi,m) \cn(\xi,m)}
{1+B\cn^2(\xi,m)}\,,~~B > 0\,,
\ee
provided the relation (\ref{3.6}) is satisfied and further if
\bea\label{3.29}
&&g = -1\,,~~B = \frac{1-\sqrt{1-m}}{\sqrt{1-m}}\,,~~0 < m < 1\,, \nonumber \\
&&\omega = (2-m-6\sqrt{1-m}) \alpha \delta^2\,,~~ A^2 
= \frac{4(1-\sqrt{1-m})^2 \beta^2}{\sqrt{1-m}}\,.
\eea

On using the identity (\ref{39}), one can re-express the solution 
(\ref{3.28}) as a 
superposition of two periodic pulse solutions, i.e.
\be\label{3.30}
u(x,t) = e^{i\omega t} \beta \big (\dn[\xi -K(m)/2,m] 
- \dn[\xi +K(m)/2,m] \big )\,.
\ee

{\bf Solution XIII}

Remarkably, the nonlocal Hirota Eq. (\ref{3.1}) also admits another periodic 
solution
\be\label{3.31}
u(x,t) = e^{i\omega t} \frac{A \dn(\xi,m)}{1+B\cn^2(\xi,m)}\,,
~~B > 0\,,
\ee
provided the relation (\ref{3.6}) is satisfied and further if
\bea\label{3.32}
&&g = 1\,,~~B = + \frac{1-\sqrt{1-m}}{\sqrt{1-m}}\,,~~0 < m < 1\,, \nonumber \\
&&\omega = [2-m+6\sqrt{1-m}] \alpha \delta^2\,,~~
A^2 = \frac{4}{\sqrt{1-m}} \beta^2\,.
\eea

On using the identity (\ref{43}), one can re-express the solution (\ref{3.31}) 
as a superposition of two periodic pulse solutions, i.e.
\be\label{3.33}
u(x,t) = e^{i\omega t} \beta  
\left(\dn[\xi +K(m)/2, m]+\dn[\xi -K(m)/2, m]\right)\,.
\ee

It is worth pointing out that while the solutions XI and XIII are also
the solutions of the local Hirota Eq. (\ref{3.2}), solution XII is only
a solution of the local Hirota Eq. (\ref{3.2}) provided $g = -1$.

\section{Conclusion and Open Problems}

In this paper we have shown that a number of nonlocal nonlinear equations
such as the Ablowitz-Musslimani variant of the nonlocal NLS \cite{abm, abm1}, 
Yang variant of the nonlocal NLS \cite{yang18}, nonlocal mKdV equation 
\cite{he18} as well as the nonlocal Hirota equation \cite{ccf,xyx} 
admit novel superposed periodic kink and pulse solutions. 
Besides, nonlocal mKdV also admits hyperbolic superposed kink solution. 
Further, we have shown that except for the Yang variant of the nonlocal
NLS, other three nonlocal equations admit both the kink and the pulse 
solutions in the same model. Besides, unlike the local nonlinear
equations the nonlocal nonlinear equations do not admit solutions with 
arbitrary translation.  However, we have
shown that all of them support solutions with definite translation. 
Finally, we have also shown that amongst these nonlocal equations, only
the Ablowitz-Musslimani variant of the nonlocal NLS admits complex
PT-invariant kink and pulse solutions. Remarkably, all these solutions
exist in the same model in which the real kink and pulse solutions 
are also admitted. 

This paper raises several
questions which are still not understood. We list some of them below.

\begin{enumerate}

\item In this paper, in several different models we have obtained a
number of solutions which can be re-expressed as 
either the sum or the difference of two $\sn(x,m)$ or two $\dn(x,m)$ 
Jacobi elliptic functions. However, we have not been able to obtain 
similar superposed solutions in the Jacobi
elliptic $\cn(x,m)$ case. It is not clear what is the underlying reason. 
Clearly it would be worthwhile finding $\cn(x,m)$ superposed solutions 
in some nonlocal nonlinear models.

\item So far only for the nonlocal mKdV equation we have been able to 
obtain a (hyperbolic) solution which can be re-expressed as the sum of 
a kink and an antikink solution. However, 
so far we have not been able to obtain hyperbolic solutions which can be 
re-expressed either as a sum of two kink or a sum or difference of two 
pulse solutions. It is clearly of interest to look for such solutions. 

\item It is not clear what is the physical interpretation of such 
superposed periodic or hyperbolic solutions. Do they correspond to a 
bound state of a kink and an antikink or of two periodic kinks or 
two periodic pulse solutions? Or do they merely correspond to some 
excitation of a kink and an antikink, or of two kink or of two pulse 
solutions? It is worthwhile finding the interpretation of
such superposed solutions vis a vis a single kink or pulse solution.

\item It is clearly of interest to discover other nonlocal nonlinear 
equations which also admit such or even more unusual superposed solutions.  

\end{enumerate}

Hopefully one can find answers to some of the questions raised above.

{\bf Acknowledgment}

One of us (AK) is grateful to Indian National Science Academy (INSA) for the
award of INSA Senior Scientist position at Savitribai Phule Pune University. 
The work at Los Alamos National Laboratory was carried out under the auspices 
of the U.S. DOE and NNSA under Contract No. DEAC52-06NA25396. 

{\bf Appendix A: Solutions of Yang's nonlocal variant of NLS Eq. (\ref{5})}

We now present those solutions of the local NLS which are also the solutions 
of the Yang's variant of the nonlocal NLS for the same value of the 
parameters.

{\bf Solution IX}

Another solution to the Yang variant of the nonlocal NLS Eq. (\ref{5}) is
\be\label{10}
\psi(x,t) = A \sqrt{m} \cn(\beta x, m) e^{i\omega t}\,,
\ee
provided
\be\label{11}
 g = 1\,,~~ A^2 = \beta^2\,,~~\omega = (2m-1)\beta^2\,.
\ee

{\bf Solution X}

Remarkably, even a linear superposition of solutions I and II is also a
solution of Eq. (\ref{5}), i.e.
\be\label{12}
\psi(x,t) = [A \dn(\beta x, m) +B \sqrt{m} \cn(\beta x, m)]
e^{i\omega t}\,,
\ee
provided
\be\label{13}
g = 1\,,~~4 A^2 = \beta^2\,,~~B = \pm A\,,~~\omega 
= \frac{(1+m)}{2} \beta^2\,.
\ee

{\bf Solution XI}

In the limit $m = 1$, the solutions I, II and III (with $B = A$) go over to
the hyperbolic solution
\be\label{14}
\psi_1(x,t) = A \sech(\beta x) e^{i\omega t}\,,
\ee
provided
\be\label{15}
 g =1\,,~~ A^2 = \beta^2\,,~~\omega = \beta^2\,,
\ee
while solution III with $B = -A$ goes over to the vacuum solution
$\psi_1 = \psi_2 = 0$.

{\bf Solution XII}

Yet another solution to the nonlocal Eq. (\ref{5}) is
\be\label{16}
\psi(x,t) = A \sqrt{m} \sn(\beta x, m) e^{i\omega t}\,,
\ee
provided
\be\label{17}
 g = -1\,,~~A^2 = -\beta^2\,,~~\omega = -(1+m)\beta^2\,.
\ee

{\bf Solution XIII}

In the limit $m = 1$, solution XII goes over to the hyperbolic solution
\be\label{18}
\psi(x,t) = A \tanh(\beta x) e^{i\omega t}\,,
\ee
provided
\be\label{19}
 g = -1\,,~~A^2 = -\beta^2\,,~~\omega = - 2 \beta^2\,.
\ee
Thus as in the local NLS case, in this nonlocal NLS model, while pulse
solutions are admitted in case $g > 0$, the kink solutions
exist only if $g < 0$. In contrast, in the Ablowitz-Musslimani 
variant of the nonlocal NLS, both the kink and pulse solutions 
exist in the same model, i.e. in case $g > 0$. 

{\bf Solution XIV: Peregrine Soliton}

Remarkably, the celebrated Peregrine soliton solution \cite{per83,dj} of 
the local NLS is also a solution of the nonlocal Eq. (\ref{5}). In 
particular, it is easy to check that 
\be\label{20}
\psi(x,t) = \frac{1}{\sqrt{2}} \bigg [1-\frac{4(1+2it)}
{(1+2x^2+4t^2)} \bigg ] e^{it}\,,~~g = 1\,,
\ee
is an exact solution of Eq. (\ref{5}). 

{\bf Solution XV: Akhmediev-Eleonskii-Kulagin Breather Solution}

Remarkably, even the celebrated Akhmediev-Eleonskii-Kulagin breather 
solution \cite{dj, akh85} of the local NLS is also a solution of the nonlocal 
Eq. (\ref{5}). In particular, it is easy to check that 
\be\label{21}
\psi(x,t) = \sqrt{\frac{a^2}{2}} e^{ia^2 t} \bigg 
[\frac{b^2 \cosh(\theta) +ib\sqrt{2-b^2}}{\sqrt{2}\cosh(\theta)-\sqrt{2-b^2}
\cos(abx)} \bigg ]\,,~~g = 1\,,
\ee
where $\theta = a^2 b \sqrt{2-b^2} t$, is an exact solution of the  
nonlocal Eq. (\ref{5}).

{\bf Solution XVI: Kuznetsov-Ma Soliton}

The celebrated Kuznetsov-Ma soliton solution \cite{dj, kuz77} 
of the local NLS is also a solution of the nonlocal Eq. (\ref{5}). 
In particular, it is easy to check that
\be\label{22}
\psi(x,t) =\frac{a}{\sqrt{2}} e^{ia^2 t}\bigg [1
+\frac{2m(m\cos(\theta) +in\sin(\theta)}{n\cosh(\sqrt{2} ma x)
+\cos(\theta)} \bigg ]\,,~~g = 1\,,
\ee
where $n^2 = 1+m^2\,,~~\theta = 2m n a^2 t$, is an exact solution of the 
nonlocal Eq. (\ref{5}). 

{\bf Appendix B: Exact Solutions of Ablowitz-Musslimani Variant of 
Nonlocal NLS Eq. (\ref{1})}

{\bf Solution XI}

In \cite{ks14} we had shown that $\dn(\beta x,m)$ as well as 
$\dn[\beta(x+K[m]),m]$ are the exact solutions of the nonlocal 
Eq. (\ref{1}). 
Remarkably, it turns out that not only $\dn[\beta x,m]$ and 
$\dn[\beta(x+K[m]),m]$ but even their superposition is a solution
of the nonlocal NLS Eq. (\ref{1}). In particular, it is easy to check
that
\be\label{1.0}
\psi(x,t) = e^{i\omega t} \left[A\dn(\beta x,m)+ \frac{B\sqrt{1-m}}
{\dn(\beta x,m)}\right]\,,
\ee
is also an exact solution of the nonlocal Eq. (\ref{5}) provided
\be\label{1.00}
g=1\,,~~ A^2 = \beta^2\,,~~B = \pm A\,,~~
\omega = [2-m \pm 6\sqrt{1-m}]\beta^2\,,
\ee
where the $\pm$ sign in $B = \pm A$ and in $\omega$ are correlated.

{\bf Solution XII: Peregrine Soliton}

It is easy to check that the celebrated Peregrine soliton solution 
\cite{per83,dj} of the local NLS is also a solution of the nonlocal 
Eq. (\ref{1}), i.e. in particular, it is easy to check that 
\be\label{1.1}
\psi(x,t) = \frac{1}{\sqrt{2}} \bigg [1-\frac{4(1+2it)}
{(1+2x^2+4t^2)} \bigg ] e^{it}\,,~~g = 1\,,
\ee
is an exact solution of Eq. (\ref{5}).

{\bf Solution XIII: Akhmediev-Eleonskii-Kulagin Breather Solution}

It is easy to check that even the celebrated Akhmediev-Eleonskii-Kulagin 
breather solution \cite{dj, akh85} of the local NLS is also a solution of 
the nonlocal Eq. (\ref{1}) 
\be\label{1.2}
\psi(x,t) = \sqrt{\frac{a^2}{2}} e^{ia^2 t} \bigg 
[\frac{b^2 \cosh(\theta) +ib\sqrt{2-b^2}}{\sqrt{2}\cosh(\theta)-\sqrt{2-b^2}
\cos(abx)} \bigg ]\,,~~g = 1\,,
\ee
where $\theta = a^2 b \sqrt{2-b^2} t$.

{\bf Solution XIV: Kuznetsov-Ma Soliton}

Remarkably, even the celebrated Kuznetsov-Ma soliton solution \cite{dj, kuz77} 
of the local NLS is also a solution of the nonlocal Eq. (\ref{1}). 
In particular, it is easy to check that
\be\label{1.3}
\psi(x,t) =\frac{a}{\sqrt{2}} e^{ia^2 t}\bigg [1
+\frac{2m(m\cos(\theta) +in\sin(\theta)}{n\cosh(\sqrt{2} ma x)
+\cos(\theta)} \bigg ]\,,~~g = 1\,,
\ee
where $n^2 = 1+m^2\,,~~\theta = 2m n a^2 t$, is an exact solution of the 
nonlocal Eq. (\ref{5}). 

{\bf Appendix C: Exact Solutions of the Nonlocal mKdV Eq. (\ref{2.1})}

We present here those solutions of the nonlocal mKdV Eq. (\ref{2.1}) 
which are also the solutions of the local mKdV Eq. (\ref{2.2}) for the
same set of parameters.

{\bf Solution XV}

Similarly, 
\be\label{2.6}
\psi(x,t) = A \sqrt{m} \cn[\beta (x-vt), m] \,,
\ee
is an exact solution to Eq. (\ref{2.1}) provided
\be\label{2.7}
g = 1\,,~~A^2 = \beta^2\,,~~v = (2m-1)\beta^2\,.
\ee

{\bf Solution XVI}

Remarkably, even a linear superposition of solutions I and II is also a
solution of Eq. (\ref{2.1}), i.e.
\be\label{2.8}
\psi(x,t) = A \dn[\beta (x-vt), m] +B \sqrt{m} \cn[\beta (x-vt), m]\,,
\ee
provided
\be\label{2.9}
g=1\,,~~ 4 A^2 = \beta^2\,,~~B = \pm A\,,~~v = \frac{(1+m)}{2} \beta^2\,.
\ee

{\bf Solution XVII}

In the limit $m = 1$, solutions I, II and III (with $B = A$) go over to
the hyperbolic solution
\be\label{2.10}
\psi(x,t) = A \sech[\beta (x-vt)]\,,
\ee
provided
\be\label{2.11}
A^2 = \beta^2\,,~~v = \beta^2\,,
\ee
while solution III with $B = -A$ goes over to the vacuum solution
$\psi = 0$.

{\bf Solution XVIII: Rational Solution}

It is easy to check that the rational solution of the nonlocal mKdV 
Eq. (\ref{2.2}) is ($g =1$) 
\be\label{2.24}
\psi(x,t) = c - \frac{4c}{4c(x-6c^2 t)^2+1}\,.
\ee

{\bf Solution XIX: The Bion Solution}

The well known breather (also called bion) solution of the attractive local 
mKdV Eq. (\ref{2.2}) is \cite{dj}
\be\label{2.20}
\psi(x,t) = -2\frac{d}{dx} \tan^{-1} \bigg [\frac{c\sin(ax+bt+a_0)}
{a\cosh(cx+dt+c_0)} \bigg ]\,,
\ee
provided
\be\label{2.21}
g = 1\,,~~b = a(a^2-3c^2)\,,~~d = c(3a^2-c^2)\,.
\ee
Here $a_0, c_0$ are arbitrary constants. It is then clear that 
the bion solution as given by Eq. (\ref{2.20}) is also the 
bion solution of the nonlocal mKdV Eq. (\ref{2.1}) provided $a_0 = c_0 = 0$.

{\bf Solution XX: Periodic Generalization of the Bion Solution}

It has been shown \cite{kks} that the periodic generalization of the bion 
solution of the local mKdV Eq. (\ref{2.2}) is
\be\label{2.22}
\psi(x,t) = -2\frac{d}{dx} \tan^{-1} \bigg [\alpha \sn(ax+bt+a_0, k)
\dn(cx+dt+c_0, m) \bigg ]\,,
\ee
provided
\bea\label{2.23}
&&a^4 k = c^4(1-m)\,,~~\alpha = \frac{c}{a}\,, \nonumber \\
&&b = a[a^2(1+k)-3c^2(2-m)]\,,~~d = c[3a^2(1+k)-(2-m)c^2]\,. ~~~
\eea
As expected, in the limit $m \rightarrow 1, k \rightarrow 0$, the periodic
bion solution (\ref{2.22}) goes over to the bion solution (\ref{2.20}) and the 
relations between $c$ and $d$ as well as between $a$ and $b$ as given by 
Eq. (\ref{2.23}) go over to the one given in Eq. (\ref{2.21}). 

It is then clear that the periodic bion solution as given by Eq. (\ref{2.22}) 
is also the periodic bion solution of the nonlocal mKdV Eq. (\ref{2.1}) 
provided $a_0 = c_0 = 0$.

{\bf Solution XXI}

It has been shown \cite{kksh} that there is another periodic solution of the
attractive local mKdV Eq. (\ref{2.2}) and it is easy to check that it is 
also the solution of the nonlocal mKdV Eq. (\ref{2.1}), and is given by
\be\label{2.25}
\psi(x,t) = -2\frac{d}{dx} \tan^{-1} \bigg [\alpha \sc(ax+bt, k)
\dn(cx+dt, m) \bigg ]\,,
\ee
provided
\bea\label{2.26}
&&g = 1\,,~~(1-k) a^{4} = (1-m) c^4\,,~~\alpha^2 = -\frac{c}{a}\,, \nonumber \\
&&b = -a[a^2(2-k)+3c^2(2-m)]\,,~~d =- c[3a^2(2-k)+(2-m)c^2]\,. ~~~~~~~
\eea

\end{document}